\documentclass{Interspeech}

\interspeechcameraready

\title{FLASepformer: Efficient Speech Separation with Gated Focused Linear Attention Transformer}

\author[]{Haoxu}{Wang}
\author[]{Yiheng}{Jiang}
\author[]{Gang}{Qiao}
\author[]{Pengteng}{Shi}
\author[]{Biao}{Tian}

\affiliation{Tongyi Lab}{Alibaba Group}{China}
\email{\{wanghaoxu.whx,jiangyiheng.jyh,songjiang.qg,pengteng.spt,tianbiao.tb\}@alibaba-inc.com}

\keywords{speech separation, focused linear attention, efficient}

\usepackage{comment}

\usepackage{amsmath,amssymb,amsfonts}
\usepackage{algorithmic}
\usepackage{graphicx}
\usepackage{textcomp}
\usepackage{multirow}
\usepackage{booktabs}
\usepackage{marvosym}
\usepackage{cite}
\usepackage{setspace} 
\usepackage{ulem}
\usepackage{makecell}
\usepackage[table]{xcolor}

\definecolor{lightgray}{gray}{0.9}

\begin{document}

\maketitle

\begin{abstract}

Speech separation always faces the challenge of handling prolonged time sequences. Past methods try to reduce sequence lengths and use the Transformer to capture global information. However, due to the quadratic time complexity of the attention module, memory usage and inference time still increase significantly with longer segments. To tackle this, we introduce Focused Linear Attention and build FLASepformer with linear complexity for efficient speech separation. Inspired by SepReformer and TF-Locoformer, we have two variants: FLA-SepReformer and FLA-TFLocoformer. We also add a new Gated module to improve performance further. Experimental results on various datasets show that FLASepformer matches state-of-the-art performance with less memory consumption and faster inference. FLA-SepReformer-T/B/L increases speed by 2.29x, 1.91x, and 1.49x, with 15.8\%, 20.9\%, and 31.9\% GPU memory usage, proving our model's effectiveness.

\end{abstract}

\section{Introduction}

Monaural speech separation (SS) extracts individual speech sources from a single-channel mixture, which is crucial for addressing the cocktail party problem\cite{cherry1953some,bregman1994auditory} and improving speech applications. 
While previous SS methods achieve good results using neural networks\cite{deepcluster,pit,sepformer}, SS still faces the challenge of modeling prolonged time sequences, resulting in high computational complexity and slow inference. Past methods use various methods to reduce sequence lengths. TF domain models use Short-Time Fourier transform (STFT) by downsampling FFT bins\cite{tfgridnet,tflocoformer}.
Some time domain models like TasNet\cite{convtasnet} use fixed receptive fields. Dual-Path Modeling, as introduced by DPRNN\cite{dprnn}, chunks the sequences for eariser training with Long-Short Term Memory (LSTM)\cite{lstm}. DPTNet\cite{dptnet} introduces Transformer\cite{attentionallyouneed} to improve long-range modeling.
In addition, some efficient speech processing models often use U-Net for downsampling and lightweight designs \cite{sudormrf,sandglasset,tdanet,tiger,zipenhancer}.

However, these methods often lose global information through STFT, fixed receptive fields, chunking, downsampling, etc. Even with downsampling, using attention mechanisms still results in quadratic complexity, causing processing time to grow as $O(N^2)$ with longer audio. On the other hand, $O(N)$ models like LSTM struggle to capture global information effectively.
Mossformer\cite{mossformer} and Mossformer2\cite{mossformer2} use GAU\cite{gau} with approximate linear time complexity to capture global information. Linear attention, on the other hand, is considered a simple and effective alternative by reducing the general complexity. Vanilla Linear Attention\cite{vla} (VLA) achieves $O(N)$ complexity by computing $K^{T}V$ first compared to $QK^{T}$ in Softmax Attention. In SS tasks $d << N$ (channel dimension is much smaller than sequence length), using an efficient linear attention scheme to model long sequences is important.

\begin{figure}[!t]
	\centering
	\includegraphics[scale=0.36]{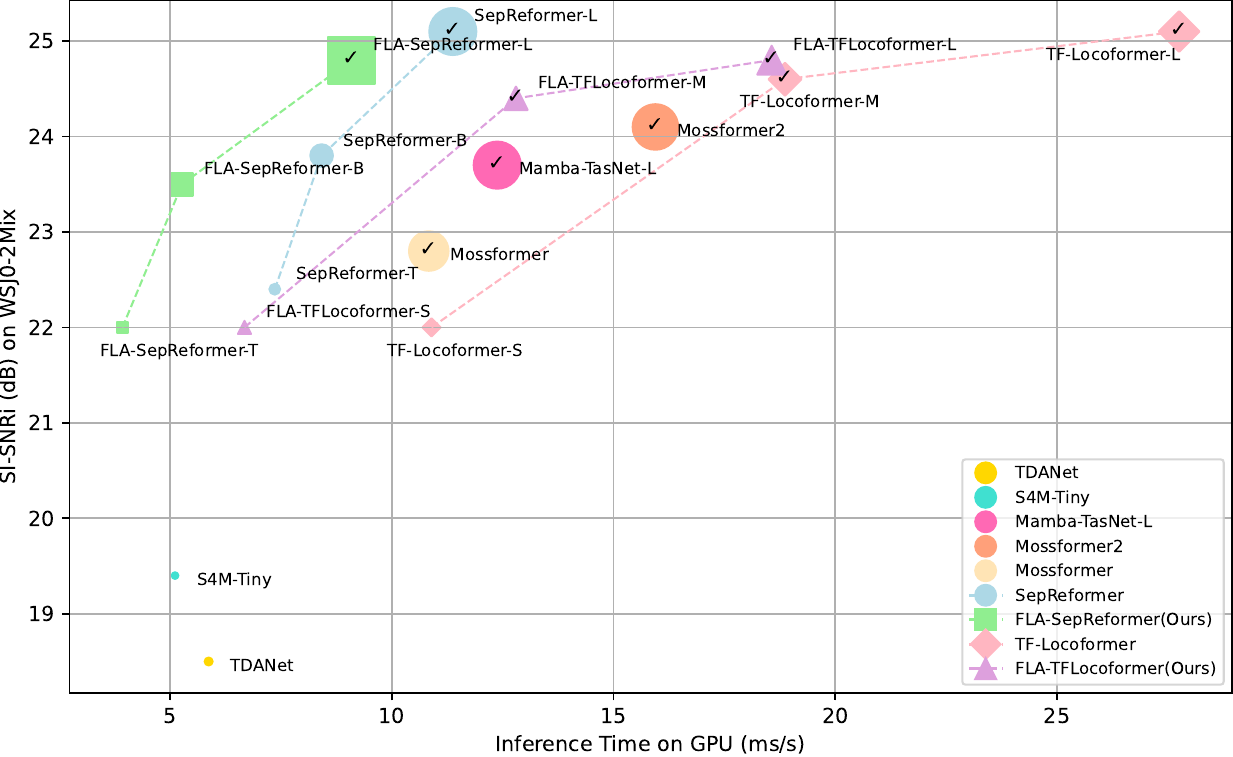}
        \vspace{-10pt}
	\caption{
		Si-SNRi results on WSJ0-2Mix versus Inference Time on RTX A800 GPU (ms/s) for 30s audio mixture. The check mark indicates using DM method for training. The radius of circle is proportional to the parameter size of the model.
	}
	\label{fig:compare}
        \vspace{-18pt}
\end{figure}


Therefore, we introduce a new Focused Linear Attention (FLA)\cite{fla} to the SS task, replacing commonly used downsampling and quadratic complexity attention mechanisms used in previous SS work\cite{sepreformer}.
FLA improves VLA by using a simple pull-push mapping method to address overly smooth attention weights. It also uses a depthwise convolution (DWC) module to alleviate the loss of feature diversity from the lower rank of VLA attention weights.
We apply FLA to the SS task, modifying the original 2D DWC module to a 1D DWC suitable for speech tasks. We also introduce a Gated module to control the output of FLA and improve model performance. Using the Gated FLA module, we build FLASepformer, primarily based on the latest models, SepReformer\cite{sepformer} and TF-locoformer\cite{tflocoformer}. We replace the MHSA w d/u in SepReformer and the attention mechanism in TF-locoformer's Temporal Modeling. 
Ultimately, we propose FLASepformer with two variants, FLA-SepReformer and FLA-TFLocoformer, achieving comparable state-of-the-art (SOTA) performance to SepReformer and TF-Locoformer on the WSJ0-2Mix dataset while greatly reducing inference complexity. 
As shown in Fig. \ref{fig:compare} and Fig. \ref{fig:inference}, FLA-SepReformer-T/B/L increases speed by 2.29x, 1.91x, and 1.49x for 30s audio mixture, with 15.8\%, 20.9\%, and 31.9\% GPU memory usage, proving our model's effectiveness. 


\begin{figure*}[!t]
	\centering
	\includegraphics[scale=0.80]{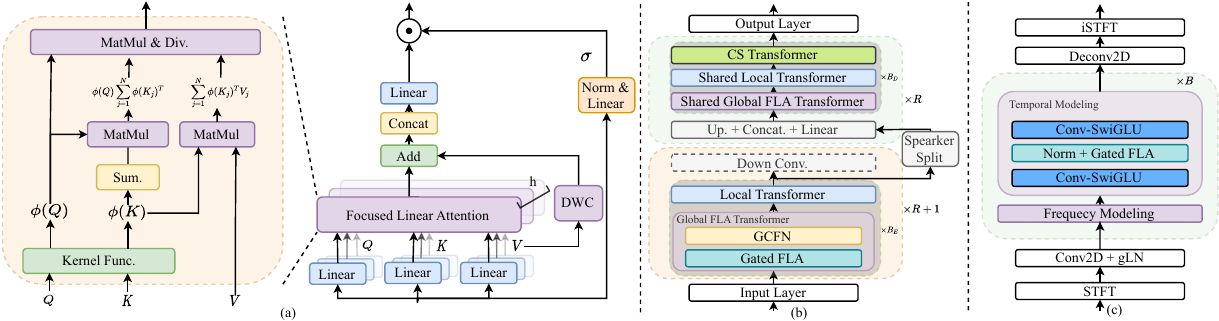}
        \vspace{-10pt}
	\caption{
        (a) The structure of Gated Focused Linear Attention Module. (b) The structure of FLA-SepReformer. (c) The structure of FLA-TFLocoformer.
	}
	\label{fig:model}
        \vspace{-15pt}
\end{figure*}



\section{Methods}

\subsection{Vanilla Linear Attention}

Let's first review the original attention module, which can be defined as: 

\begin{equation}
\begin{array}{cc}
    Q = XW^{q}, K = XW^{k}, V = XW^{v}, \\
    O_{i} = \sum^{N}_{j=1}\frac{Sim(Q_{i}, K_{j})}{\sum^{N}_{j=1}Sim(Q_{i}, K_{j})}V_{j} \\
\end{array}
\end{equation}


\noindent which $X \in R^{N \times d}$ is the input sequence, $W^{*} \in R^{d \times d}$ is the linear layer, and $i$ is the sequence index. Traditional attention uses Softmax attention as the similarity function $Sim(Q, K) = exp(QK^{T}/sqrt(d))$ with $O(N^2)$ complexity, which is inefficient for long sequences in speech separation. Linear attention adjusts the similarity calculation to $Sim(Q, K) = \phi(Q)\phi(K)^{T}$, where $\phi(\cdot)$ is a kernel function, providing a more efficient alternative. The linear attention module is defined as:

\begin{equation}
\begin{array}{cc}
    O_{i} = \sum^{N}_{j=1}\frac{\phi(Q)\phi(K)^{T}}{\sum^{N}_{j=1}\phi(Q)\phi(K)^{T}}V_{j}
\end{array}
\end{equation}


\noindent by adjusting the similarity function, we can first compute $\phi(K)^{T} V$ using matrix multiplication law, defined as:

\begin{equation}
\begin{array}{cc}
    O_{i} = \frac{\phi(Q)(\sum^{N}_{j=1}\phi(K_{j})^{T}V_{j})}{\phi(Q)\sum^{N}_{j=1}\phi(K)^{T}}
\end{array}
\end{equation}

\noindent which reduces the computational complexity from $O(N^2)$ to $O(N)$, allowing for parallel training and linear inference time compared to Softmax Attention.



However, current linear attention mechanisms face a trade-off between accuracy and complexity. Simple kernels like RELU fail to model long-range relationships between $Q$ and $K$, leading to smooth attention weights and degrading performance\cite{efficientviT}. Complex kernel functions may increase computational complexity\cite{Complexkf}. Additionally, in the SS task, VLA may not effectively capture speech feature diversity in very long sequences.

\subsection{Gated Focused Linear Attention}
\label{sec:GFLA}


The whole architecture of Gated Focused Linear Attention is shown in Fig.\ref{fig:model}(a). We introduce Focused Linear Attention (FLA) into the SS task to better model long speech sequence features. FLA improves smooth attention weights from VLA. FLA introduces a novel kernel function, Focused Function $\phi_{p}(\cdot)$, to enhance the model's focus ability on different features. This helps effectively combine important features from long sequences for various queries. The new kernel function, Focused Function $\phi_{p}(\cdot)$, is defined as:

\begin{equation}
\begin{array}{cc}
    \phi_{p}(x) = f_{p}(RELU(x)), f_{p}=\frac{||x||}{||x^{**p}||}x^{**p}
\end{array}
\end{equation}


\noindent where $x^{**p}$ denotes the element-wise power $p$ of $x$. By designing an appropriate focus factor $p$, FLA pulls similar query-key pairs closer and pushes dissimilar pairs apart. This design enhances the aggregation of similar speech token features while reducing the gathering of unrelated ones. Consequently, FLA improves VLA to mimic the sharp distribution of Softmax Attention weights, effectively modeling important long-range speech features.

Compared to Softmax Attention, VLA needs to build both sharper attention weights and feature weights with high diversity. According to \cite{fla}, the maximum rank of VLA's attention weights depends on the number of speech features $N$ and the channel dimension $d$, given by $rank(\phi(Q)\phi(K)^{T}) \leq min\{rank(\phi(Q)), rank(\phi(K))\} \leq min\{N, d\}$. In SS tasks, $d$ is often much smaller than $N$; for instance, in SepReformer-B\cite{sepreformer}, $d=16$ and $N=2000$ for 1 second of audio.
This leads to significantly less feature diversity in VLA than in Softmax Attention. The original FLA uses a 2D depthwise convolution (DWC2d) module to enhance attention weight rank, defined as $O = \phi(Q)\phi(K)^{T}V + DWC2d(V)$. While DWC2d targets local information in images, for speech separation, we switch it to DWC1d with kernel size $k$ for one-dimensional speech features. The DWC1d module improves attention weight diversity and, with a full-rank attention matrix similar to Softmax Attention, compensates for the rank lost in $\phi(Q)\phi(K)^{T}$, matching the original upper bound. Additionally, it also acts as a local attention mechanism, enhancing information about adjacent features in different heads within the time domain.

Compared to the VLA, FLA modifies only the kernel function to the Focused Function and adds a fixed-cost DWC1d module, maintaining $O(N)$ complexity. This enables the modeling of long-range speech features and builds an efficient speech separation model while reducing inference complexity to linear compared to the $O(N^2)$ complexity of Softmax Attention.

To enhance FLA's modeling capability further, we introduce a gating mechanism inspired by \cite{gau}. We use a Gated MultiLayer Perceptron (MLP), consisting of a Norm layer, a linear layer, and an activation function to obtain the gating result, which is then multiplied by FLA's output. This Gated FLA module uses current token features to control global feature interaction better.

\subsection{The architecture of the FLASepformer}

\begin{table}[t]\centering
\small
\setlength{\tabcolsep}{0.2mm}{
  \centering
  \footnotesize
  \caption{\label{tab:wsj02mix} {Comparisons with other methods on WSJ0-2Mix. Results in [dB]. $*$ represents the results of our reproduced model.
  \vspace{-7pt}
}}
  \begin{tabular}{cccccccccccc}
  \toprule
  \multirow{2}*{\textbf{Methods}} & \multirow{2}*{\makecell[c]{\textbf{Param} \\ \textbf{[M]}} } & \multirow{2}*{\makecell[c]{\textbf{MACs} \\ \textbf{[G]}} }  & \multicolumn{2}{c}{\textbf{w/o DM}} & \multicolumn{2}{c}{\textbf{w/ DM}}  \\
  \cmidrule(lr){4-5} \cmidrule(lr){6-7} & & & SI-SNRi & SDRi & SI-SNRi & SDRi \\
  
  \midrule
  
  Conv-TasNet\cite{convtasnet} & 5.1 & 5.1 & 15.3 & 15.6 & - & - \\
  SuDoRM-RF\cite{sudormrf} & 6.4 & - & 18.9 & - & - & - \\
  TDANet\cite{tdanet} & 2.3& -& 18.5& 18.7& -& -\\
  Sandglasset\cite{sandglasset} & 2.3& -& 20.8& 21.0& -& -\\
  S4M-Tiny\cite{s4m} & 1.8 & -& 19.4& 19.7& -& -\\\
  S4M\cite{s4m} & 3.6 & -& 20.5& 20.7& -& -\\

  \midrule
  DPRNN\cite{dprnn} & 2.6 & 42.2 & 18.8 & 19.0 & - & - \\
  DPTNet\cite{dptnet} & 2.7 & - & 20.2 & 20.6 & - & -  \\
  SepFormer\cite{sepformer} & 25.7 & 59.5 & 20.4 & 20.5 & 22.3 & 22.4 \\
  TF-GridNet\cite{tfgridnet} & 14.4 & 231.1 & 23.5 & 23.6 & - & - \\
  Mamba-TasNet(L)\cite{mambatasnet} & 59.6 & - & - & - & 23.7 & 23.8 \\
  MossFormer\cite{mossformer} & 42.1 & 42.7$^{*}$ & - & - & 22.8 & - \\
  MossFormer2\cite{mossformer2} & 55.7 & 56.4$^{*}$ & - & - & 24.1 & - \\

  \midrule

  SepReformer-T\cite{sepreformer} & 3.7 & 5.2$^{*}$ & 22.4 & 22.6 & - & - \\
  SepReformer-B\cite{sepreformer} & 14.2 & 19.9$^{*}$ & 23.8 & 23.9 & - & - \\
  SepReformer-L\cite{sepreformer} & 59.4 & 77.7$^{*}$ & - & - & 25.1 & 25.2 \\
  SepReformer-B (Rep.) & 14.2 & 19.9$^{*}$ & 23.6 & 23.7 & - & - \\
  \rowcolor{lightgray}
  FLA-SepReformer-T & 3.7 & 5.6 & 22.0 & 22.1 & - & - \\
  \rowcolor{lightgray}
  FLA-SepReformer-B & 14.2 & 21.6 & 23.5 & 23.7 & - & - \\
  \rowcolor{lightgray}
  FLA-SepReformer-L & 59.4 & 84.6 & - & - & 24.7 & 24.8 \\

  \midrule
  TF-Locoformer-S\cite{tflocoformer} & 5.0 & 43.7 & 22.0 & 22.1 & 22.8 & 23.0 \\
  TF-Locoformer-M\cite{tflocoformer} & 15.0 & 127.8 & 23.6 & 23.8 & 24.6 & 24.7 \\
  TF-Locoformer-L\cite{tflocoformer} & 22.5 & 191.7 & 24.2 & 24.3 & 25.1 & 25.2 \\
  \rowcolor{lightgray} 
  FLA-TFLocoformer-S & 5.2 & 44.0 & 22.1 & 22.3 & 22.8 & 22.9 \\
  \rowcolor{lightgray}
  FLA-TFLocoformer-M & 15.1 & 128.7 & 23.4 & 23.5 & 24.4 & 24.5 \\
  \rowcolor{lightgray}
  FLA-TFLocoformer-L & 22.6 & 193.0 & 24.2 & 24.3 & 24.8 & 24.9 \\

  \bottomrule

\end{tabular}
}
\vspace{-18pt}
\end{table}

We build FLASepformer, with two variants: FLA-SepReformer and FLA-TFLocoformer, based on SepReformer\cite{sepformer} and TF-Locoformer\cite{tflocoformer}, respectively. In Fig.\ref{fig:model}(b), for SepReformer, we replace the original efficient global attention (GLA) in all global transformers with the Gated FLA from Sec. \ref{sec:GFLA}. Unlike GLA, which uses downsampled MHSA to reduce the sequence length to $\frac{N}{2^{R}}$ while still maintaining $O(N^2)$ complexity, Gated FLA can model long-range features without downsampling and reduce the complexity to $O(N)$. Other components like the Separation Encoder, Reconstruction Decoder, Local Transformer, CS Transformer, and Speaker Split remain unchanged.

In Fig.\ref{fig:model}(c), for TF-Locoformer, we replace the original MHSA in Temporal modeling with Gated FLA. Although STFT helps reduce the length of speech features, the original MHSA maintains $O(N^2)$ complexity. Using Gated FLA, we convert quadratic complexity to linear, improving inference efficiency. Other components like Frequency Modeling and Conv-SwiGLU remain unchanged.

\section{Experimental Setup}

\subsection{Dataset}

We validate our model's performance using four popular speech separation datasets: WSJ0-2Mix\cite{deepcluster}, WHAM!\cite{wham}, WHAMR!\cite{whamr}, and Libri2Mix\cite{cosentino2020librimix}. We train and test all datasets using the full overlap min version with a sampling rate of 8kHz.

\noindent \textbf{WSJ0-2Mix} is a commonly used benchmark for SS, created from the WSJ0 corpus to generate 2-speaker clean mixtures. It consists of 30h, 10h, and 5h training, validation and test data.


\noindent \textbf{WHAM!/WHAMR!} are the noisy and noisy-reverberant versions of WSJ0-2Mix. WHAM! introduces noise from urban environments, mixed with speech at SNRs between -6 and +3 dB. WHAMR! further adds reverberation to the clean sources in WHAM!, which is used to train models for dereverberation, denoising, and speech separation.

\noindent \textbf{Libri2Mix} includes two training sets: train-360 with 106 hours and train-100 with 29 hours. Source speakers are drawn from LibriSpeech\cite{librispeech} sets (train-100, train-360). Both validation and test sets are 5.5 hours each. For a fair comparison with previous work, we train FLA-SepReformer models on train-100 and FLA-TFLocoformer models on train-360.

\subsection{Training and Model Configuration}

We use the ESPnet-SE\cite{lu22c_interspeech} toolkits for all experiments. For FLA-SepReformer, we mainly study the T/B/L model scale with settings similar to SepReformer: up to 200 training epochs, batch size of 2, and an initial learning rate (LR) of 1e-3 with a 1k-step warm-up. The LR is fixed for the first 50 epochs and then decays 0.8 if validation loss doesn't improve for 2 epochs. We use the same multi-loss as SepReformer. LR adjustments for dynamic mixing (DM) follow \cite{sepformer}.


For FLA-TFLocoformer, we mainly study the S/M/L model scale, with settings similar to TF-Locoformer: up to 150 training epochs, batch size of 4, and an initial LR of 1e-3 with a 4k-step warm-up. The LR is halved if validation loss doesn't improve for 3 epochs. LR for S model remains unchanged for 50 epochs. With DM, training extends to 200 epochs, and the LR remains unchanged for 75/75/65 epochs for S/M/L, respectively.

We set FLA's focused factor $p$ to 3 and the DWC1d kernel size $k$ to 7. Gated MLP uses LayerNorm for SepReformer and RMSGroupNorm for TF-Locoformer. Both models utilize the AdamW optimizer with a 0.01 weight decay. Audio length is set to 4 seconds for training, and gradients are clipped at an L2 norm of 5. We use speed perturbation when doing DM. All experiments are conducted on a GeForce RTX A800. We  plan to release the source code at a later time.

\begin{figure*}[!t]
	\centering
	\includegraphics[scale=0.75]{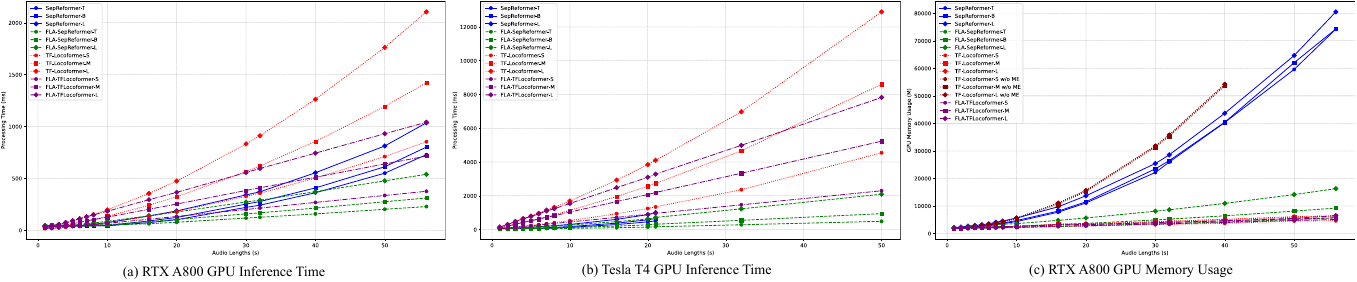}
        \vspace{-15pt}
	\caption{
    Inference Time or Memory Usage using various GPU.
	}
	\label{fig:inference}
        \vspace{-18pt}
\end{figure*}

\section{Results and Discussion}

\begin{table}[t]\centering
\small
\setlength{\tabcolsep}{0.8mm}{
  \centering
  \footnotesize
  \caption{\label{tab:whamwhamr} {Comparisons with others methods on various dataset. DM is not used on all Libri2Mix-100 results. Results in [dB]. 
    \vspace{-7pt}
}}
  \begin{tabular}{ccccccc}
  \toprule
  \multirow{2}*{\textbf{Methods}} & \textbf{WHAM!} & \textbf{WHAMR!} & \textbf{Libri2Mix-100} \\ 
  \cmidrule(lr){2-2} \cmidrule(lr){3-3} \cmidrule(lr){4-4} & \makecell[c]{SI-SNRi \\ /SDRi} & \makecell[c]{SI-SNRi \\ /SDRi} & \makecell[c]{SI-SNRi \\ /SDRi} \\
  
  \midrule
  
  Conv-TasNet\cite{convtasnet} & 12.7/- & 8.3/- & 12.2/12.7 \\ 
  SuDoRM-RF\cite{sudormrf} & 13.7/14.1 & -/- & 14.0/14.4 \\ 
  TDANet\cite{tdanet} & 15.2/15.4 & -/- & 17.4/17.9 \\ 
  S4M-Tiny\cite{s4m} & -/- & -/- & 16.2/16.6 \\ 
  S4M\cite{s4m} & -/- & -/- & 16.9/17.4 \\ 
  TF-GridNet\cite{tfgridnet} & -/- & 17.1/15.6 & -/- \\ 
  Sepformer + DM\cite{sepformer} & 16.4/16.7 & 14.0/13.0 & -/- \\ 
  MossFormer(L) + DM\cite{mossformer} & 17.3/- & 16.3/- & -/- \\ 
  MossFormer2 + DM\cite{mossformer2} & 18.1/- & 17.0/- & -/- \\ 

  \midrule
  SepReformer-T & 17.2/17.5 & -/- & 19.7/20.2 \\ 
  SepReformer-B & 17.6/18.0 & -/- & 21.7/22.1 \\ 
  SepReformer-L + DM & 18.5/18.7 & 17.1/16.0 & -/- \\ 
  \rowcolor{lightgray}
  FLA-SepReformer-T & 16.9/17.2 & 14.8/13.6 & 19.1/19.5 \\ 
  \rowcolor{lightgray}
  FLA-SepReformer-B & 17.4/17.8 & 15.6/14.4 & 20.3/20.7 \\ 
  \rowcolor{lightgray}
  FLA-SepReformer-L + DM & 18.1/18.5 & 16.4/15.2 & -/- \\ 
  \midrule
  TF-Locoformer-S & -/- & 17.4/15.9 & -/- \\ 
  TF-Locoformer-M & -/- & 18.5/16.9 & -/- \\ 
  \rowcolor{lightgray}
  FLA-TFLocoformer-S & -/- & 17.7/16.0 & -/- \\ 
  \rowcolor{lightgray}
  FLA-TFLocoformer-M & 17.5/17.7 & 18.7/17.0 &  -/- \\ 

  \bottomrule

\end{tabular}
}
\vspace{-10pt}
\end{table}

\begin{table}[t]\centering
\small
\setlength{\tabcolsep}{1.35mm}{
  \centering
  \footnotesize
  \caption{\label{tab:librimix360} {Comparisons with other methods on Libri2Mix-360. DM is not used. Results in [dB]. 
    \vspace{-7pt}
}}
  \begin{tabular}{ccccccc}
  \toprule
  \textbf{Methods} & SI-SNRi & SDRi \\
  
  \midrule
  
  Conv-TasNet\cite{convtasnet} & 14.7 & -  \\
  Sepformer\cite{sepformer} & 19.2 & 19.4 \\
  MossFormer2\cite{mossformer2} & 21.7 & - \\
  TF-Locoformer-M\cite{tflocoformer} & 22.1 & 22.2 \\
  \rowcolor{lightgray}
  FLA-TFLocoformer-M & 22.2 & 22.4 \\

  \bottomrule

\end{tabular}
}
\vspace{-7pt}
\end{table}

\begin{table}[t]\centering
\small
\setlength{\tabcolsep}{1.35mm}{
  \centering
  \footnotesize
  \caption{\label{tab:ablation} {Ablation study on WSJ0-2Mix using FLA-SepReformer-B. Results in [dB].
    \vspace{-7pt}
}}
  \begin{tabular}{ccccccc}
  \toprule
  \textbf{Gated} & \textbf{$p$} & \textbf{$k$} & SI-SNRi & SDRi \\
  
  \midrule
  
  $\checkmark$ & 3 & 7 & 23.5 & 23.7  \\
   & 3 & 7 & 23.4 & 23.5  \\
  $\checkmark$ & 2/4 & 7 & 23.4/23.3 & 23.6/23.4  \\
  $\checkmark$ & 8/16 & 7 & 23.1/22.9 & 23.3/23.0  \\
  $\checkmark$ & 3 & 15/65 & 23.4/23.4 & 23.6/23.6  \\

  \bottomrule

\end{tabular}
}
\vspace{-18pt}
\end{table}

\subsection{Comparison with previous models}

We use SI-SNR improvement (SI-SNRi) and SDR improvement (SDRi)\cite{sisdr} to evaluate model performance. Also, we report the number of multiply-accumulate operations (MACs) for 8k samples using pytorch-OpCounter\footnote{https://github.com/Lyken17/pytorch-OpCounter}. Table \ref{tab:wsj02mix} shows results on the clean WSJ0-2Mix dataset, divided into four sections: efficient SS Model, normal SS Model, SepReformer/FLA-SepReformer, and TF-Locoformer/FLA-TFLocoformer. Our FLA-SepReformer-T surpasses previous efficient SS models like TDANet and S4M. Despite slight reductions on SI-SNRi and SDRi, it matches SepReformer in parameter count while reducing computational complexity from $O(N^2)$ to $O(N)$, leading to much faster inference and lower memory usage for long sequences, as shown in Figure \ref{fig:inference}.


Additionally, we reproduce SepReformer-B (Rep.), and our FLA-SepReformer-B achieves similar results. FLA-TF-Locoformer also matches TF-Locoformer's performance across all sizes (S/M/L) while having faster inference and lower memory usage for long sequences, proving our model's efficiency. Our largest models, FLA-SepReformer-L and FLA-TFLocoformer-L, outperform prior SOTA models like TF-GridNet, MossFormer2, and those using Mamba mechanisms. This shows our models' capability to deliver faster inference with enhanced performance, highlighting the effectiveness of the Gated FLA module.

Tables \ref{tab:whamwhamr} and \ref{tab:librimix360} show results on the WHAM!, WHAMR!, and Libri2Mix-100/360 datasets, demonstrating the robust generalization of FLA-SepReformer and FLA-TFLocoformer in noisy, reverberant conditions. Our models achieve performance similar to SepReformer and TF-Locoformer, even better in some metrics, such as Libri2Mix-360 FLA-TFLocoformer-M with an SI-SNRi of 22.2.

\subsection{Inference time}

Our key contribution is matching the results of SepReformer and TF-Locoformer while reducing the attention mechanism's complexity from quadratic to linear for long sequence SS. This ensures linear growth in inference time and memory usage, preventing quadratic increases for lengthy speech. As Figure \ref{fig:compare} shows, for 30s speech, FLA-SepReformer-T/B/L is faster by 2.29x/1.91x/1.49x than SepReformer-T/B/L, and FLA-TFLocoformer-S/M/L is faster compared to TF-Locoformer-S/M/L. Additionally, FLA-SepReformer-T surpasses other efficient SS Models.
We discover inaccuracies MACs for recent models using Pytorch-OpCounter and don't align to the inference time. We analyze GPU inference time and memory usage for FLA-SepReformer/TFLocoformer, observing linear growth in inference time and memory for FLA-SepReformer versus SepReformer. The test environment for GPU are RTX A800
, Tesla T4, single-threaded. SepReformer-L meets Out-of-Memory at 57s on an 80GB setup, while models with linear complexity achieve better efficiency for long speech. Though TF-Locoformer cuts memory usage to $O(\sqrt{N})$ with Memory-Efficient (ME) Attention \cite{osqrtmem}, its inference time grows quadratically $O(N^2)$. And without ME, memory usage grows quadratically. In contrast, FLA-TFLocoformer maintains stable linear growth. Inference time calculations are averages from 100 iterations for each audio length.

\subsection{Ablation Study}


Table \ref{tab:ablation} shows the FLA-SepReformer-B results with various module and parameter modifications. Removing the Gated MLP drops SI-SNRi from 23.5 to 23.4 and SDRi from 23.7 to 23.5. Testing different focused factor $p$ values reveals that a lower value ($p=2$) reduces SI-SNRi to 23.4, and higher $p$ values also decrease performance. Adjusting the DWC1d kernel size $k$ shows that larger sizes (like $k=15/65$, similar to LocalTransformer) does not improve results. These results highlight the Gated MLP's role in gating global features and suggest that optimal $p$ and $k$ values enhance performance.

\section{Conclusion}


In this paper, we build FLASepformer, an efficient speech separation model with linear complexity. Although previous methods use STFT and downsampling to reduce speech sequence length, the attention module in those still has $O(N^2)$ time complexity. We improve SepReformer and TF-Locoformer by integrating Focused Linear Attention, creating two variants: FLA-SepReformer and FLA-TFLocoformer. We also add a new Gated module to improve performance. Experimental results on various datasets show that our models achieve similar SOTA results with reduced memory consumption and improved inference speed. FLA-SepReformer achieves speedups of 2.29x/1.91x/1.49x, and FLA-TFLocoformer-L also demonstrates significant speed gains. In the future, we will explore scenarios with more speakers.

\begin{spacing}{0.95}
\footnotesize
\normalem
\bibliographystyle{IEEEtran}
\bibliography{mybib}
\end{spacing}

\end{document}